\documentclass[twocolumn,preprintnumbers,amsmath,amssymb]{revtex4-1}

\usepackage{graphicx}
\usepackage{dcolumn}
\usepackage{bm}
\usepackage{color}
\usepackage{ulem}

\begin{document}

\title{Soliton attenuation and emergent hydrodynamics in fragile matter}
\author{N. Upadhyaya$^{\dag}$}
\author{ L. R. G\'omez$^{\dag\dag}$}
\author{ V. Vitelli$^{\dag}$}
\email{vitelli@lorentz.leidenuniv.nl}

\affiliation{$^{\dag}$ Instituut-Lorentz for Theoretical Physics, Universiteit Leiden, 2300 RA Leiden, The Netherlands \\
$^{\dag\dag}$ Department of Physics, Universidad Nacional del Sur- IFISUR - CONICET, 8000 Bah\'ia Blanca, Argentina}

\date{\today}

\begin{abstract}
Disordered packings of soft grains are fragile mechanical systems that
loose rigidity upon lowering the external pressure towards zero.
At zero pressure, we find that any infinitesimal strain-impulse propagates initially as a
non-linear solitary wave progressively attenuated by disorder. We
demonstrate that the particle fluctuations generated by
the solitary-wave decay,  can be viewed as a granular analogue of
temperature. Their presence is manifested by two emergent macroscopic properties
absent in the unperturbed granular packing: a finite pressure that scales with
the injected energy (akin to a granular temperature) and an
anomalous viscosity that arises even when the microscopic mechanisms
of energy dissipation are negligible.  Consistent with the interpretation of this
state as a fluid-like thermalized state, the shear modulus remains
zero.  Further, we follow in detail  the attenuation of the
initial solitary wave identifying two distinct regimes : an
initial exponential decay, followed by a longer power law decay
and suggest simple models to explain these two regimes.
\end{abstract}

\maketitle

\section{Introduction}

The defining property of solid materials is their ability to
resist external mechanical perturbations, as embodied in their
finite elastic moduli. However, it is the geometry and topology of
a materials' architecture that ultimately controls the strength of
the elastic response. For example, if we remodel a solid block of
steel into a granular aggregate of steel balls just in contact
with their nearest neighbours, both the linear elastic moduli and
sound speed of the aggregate drop to zero. These extreme softness,
that we term fragility, stems directly from the {\it anharmonic}
interaction between macroscopic grains in contact. It is
independent of the material the grain is made of which could be as
hard as steel or as soft as bubbles. Weakly connected {\it
harmonic} networks of cross-linked polymers or covalent glasses
can also become fragile, irrespective of the individual bond
strength, provided that the number of mechanical constraints is
too low to maintain rigidity.

The critical point at which a packing of soft repulsive grains
become fragile (unjams) can be accessed experimentally by progressively
reducing the pressure, or alternatively the average overlap, to
zero. Right at the critical point, even the tiniest mechanical
strains must propagate in the form of supersonic solitons and
shocks since the linear sound speed is zero. Nesterenko coined the
term sonic vacuum to designate this state in a granular chain and
demonstrated the existence of solitonic excitations both
experimentally and theoretically \cite{Nesterenko_Book,Nesterenko_1984}. The fate of these strongly
non-linear excitations in amorphous two or three dimensional
granular media near the (un)jamming transition or in weakly
connected polymer networks is only gradually emerging \cite{Gomez_2012}.

The very existence of these non-linear excitations and the
associated critical points impinges on a very precise tuning of
some geometrical parameters like the particle overlap that must
vanish at random close-packing or the coordination number of the
network that must be tuned exactly to maintain rigidity. It is
natural to enquire how fluctuations (that we expect to be violent
given the absence of linear restoring forces) around these special
points change their critical behavior. However, the study of
fragile nano-mechanical systems affected by quantum fluctuations
is still in its infancy. Even the impact of thermal fluctuations
on the un-jamming transition is not very well studied, partly
because granular media, that represent the key experimental arena
for these investigations, are clearly athermal.


\begin{figure}
\begin{center}
\includegraphics[width=0.45 \textwidth]{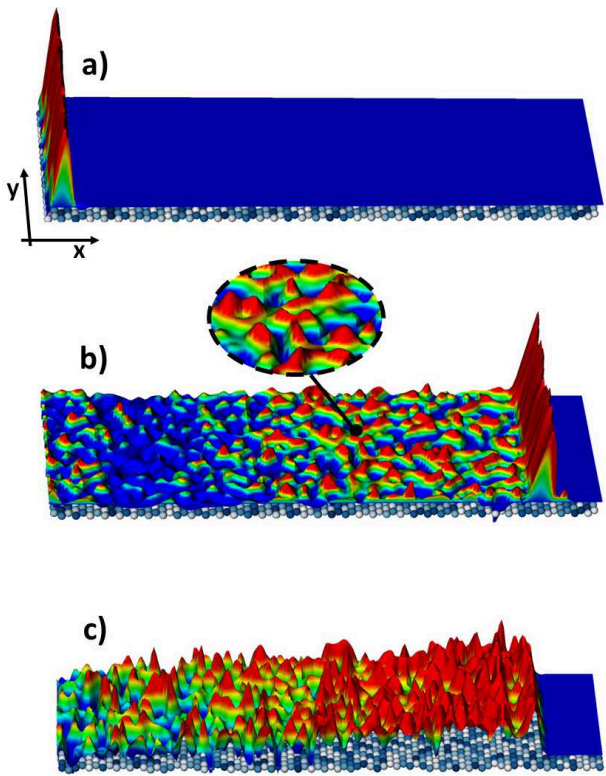}
\caption{\label{Schematic} (a) A schematic illustration of how a
hexagonal packing at zero pressure (with vanishing longitudinal
and transverse sound speeds) and weak mass disorder, responds to
an impact at one of its ends, by generating a solitary wave
excitation. (b) As the solitary wave propagates and interacts with
the weak mass inhomogeneity, its amplitude decays. The decay at
early times is nearly exponential and the process of energy loss
is such, that it excites several smaller solitary waves. Over the
long run, a noisy state (velocity fluctuations) spanning the size
of the packing emerges (zoomed region). (c) The initial solitary
wave is now no longer identifiable.  Instead we observe a
triangular shock like profile that is defined as an envelope over
the smaller excitations. In this regime, the decay of the leading
edge follows a power law.  Eventually, no leading propagating edge
is visible.}
\end{center}
\end{figure}

In this article, we study numerically and analytically the decay of solitary wave excitations in two dimensional amorphous or mass-disordered packings of grains {that are} just in contact with their nearest neighbors, see Fig.\
(\ref{Schematic} a-b). For weak mass disorder, the solitary
wave excitation generated in response to an impulse, decays
exponentially at early times, with a rate that depends upon the
amount of disorder. In the long time limit, the initially well defined solitary wave soon transitions into a
triangular shock like profile, decaying as a power law, with a universal exponent consistent with $\frac{1}{2}$ and 
independent of the amount of disorder. This power law decay is the dominant
mechanism of attenuation in hexagonal lattices with strong mass disorder as well as in jammed packings.
Finally, we study the fluctuating state (similar to a
thermalized state) that emerges after the complete disintegration
of the initial solitary wave, see the inset of {Fig.\
(\ref{Schematic}b)} and the trail of the soliton in {Fig.\
(\ref{Schematic}c)}. The resulting fluid-like state has two
emergent macroscopic properties absent in the unperturbed granular
packing: (i) a finite pressure that scales with the energy
originally injected in the form of solitary wave  and (ii) a
finite viscosity, that arises despite no microscopic dissipative mechanism are present in our simulations.

\begin{center}
\begin{figure}
\includegraphics[width=0.4\textwidth]{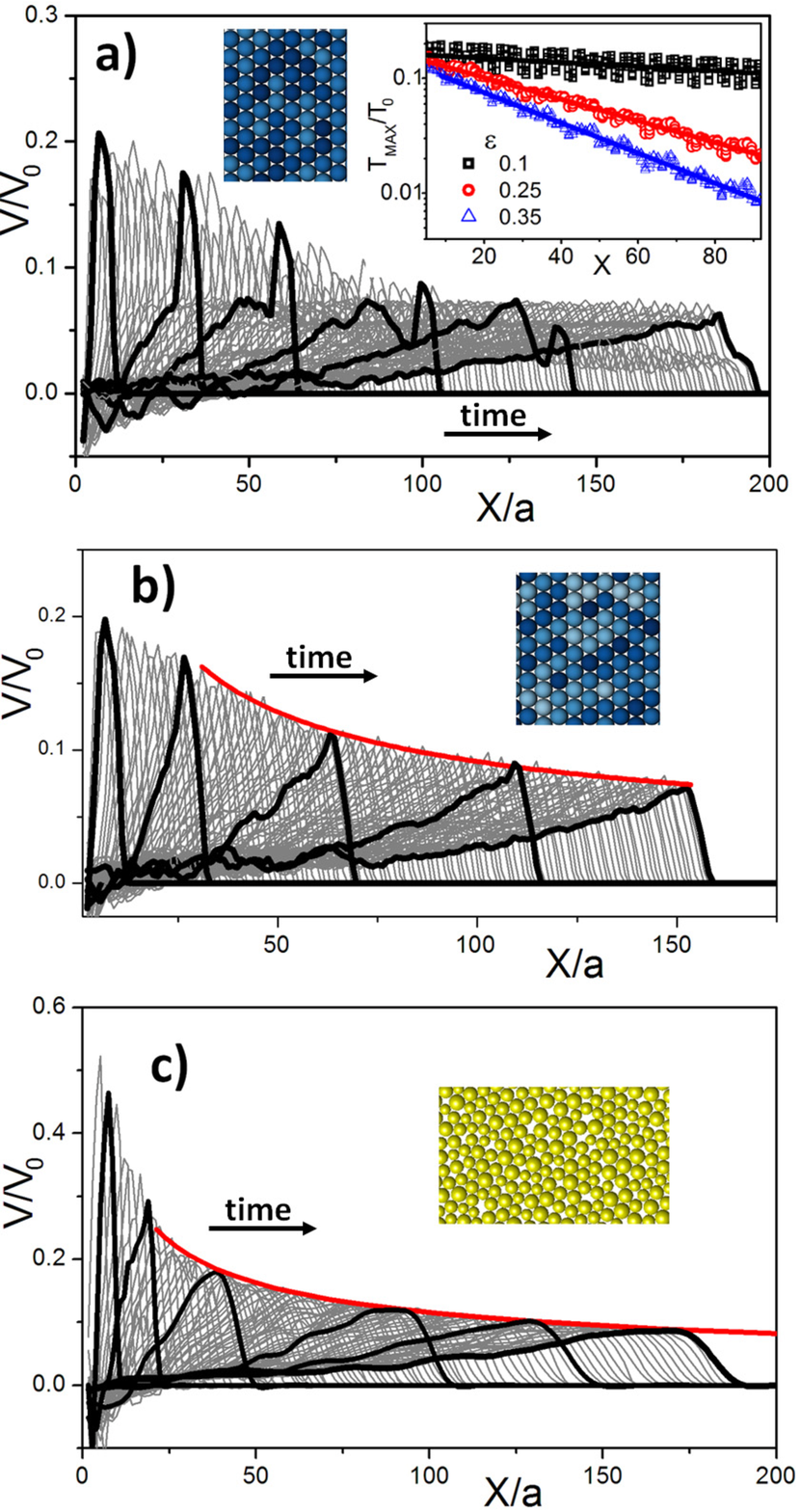}
\caption{\label{1D}(a): The time evolution in response to an
impulse  in a hexagonal packing with weak mass disorder
$\epsilon$. An impulse response generates a solitary wave
excitation that decays exponentially at early times. In the long
time limit, a shock like triangular profile emerges whose leading
edge decays as a power law with an exponent $x^{-0.5}$ (solid red
line). (b)The time evolution in response to an impulse  in a
hexagonal packing with strong mass disorder $\epsilon$. An impulse
response generates a solitary wave excitation but the regime of
exponential attenuation is barely identifiable.  Instead,  a shock
like triangular profile soon emerges whose leading edge again
decays as a power law with an exponent $x^{-0.5}$ (solid red
line). (c) The time evolution in response to an impulse  in a
jammed amorphous packing prepared at a pressure of P$\sim 10^{-6}$
and an average overlap between grains $\delta_0\sim 10^{-4}$. An
impulse response still generates a solitary wave excitation but
like (b),  a shock like triangular profile soon emerges whose
leading edge decays as a power law with an exponent $x^{-0.5}$
(solid red line). }
\end{figure}
\end{center}

\section{Attenuation of non-linear solitary waves}

To study the dynamical response near the critical point, we
perform molecular dynamics simulations for $N=4000$ soft
frictionless particles arranged in a hexagonal lattice or
as a jammed amorphous packing. The particles interact pair-wise
with the Hertz potential $V(\delta) =
K\left(\frac{\delta}{2R}\right)^{5/2}$, where $K$ sets the energy
scale and the particle overlap $\delta$ is measured in units of
the average disk radii $R$. By discretizing the packings into bins
along the $x-$direction (defined as the longitudinal direction
here), we impart an initial speed $U_p$ to the left most bin and
obtain the subsequent dynamics by integrating the equations of
motion using the velocity Verlet method \cite{Allen_Book}. We
follow how the initial energy imparted to the packing evolves into
a wave profile that propagates along the direction of initial
impact (see schematic Fig.\ (\ref{Schematic})) by averaging out
the transverse speeds over the bins.

In Fig.\ ( \ref{1D}), we show the time evolution of the solitary
wave excitation generated in response to an impulse for three
cases: (a) a hexagonal packing with grain masses distributed as a
normal random variable with a small variance $\epsilon$ (b) the
same hexagonal packing with large variance in the masses of the
grains  and (c) a jammed amorphous packing prepared at a
vanishingly small pressure P$\sim 10^{-6}$. In all three cases, we
find that in response to an impulse, the mechanical excitation
initially evolves into a well defined solitary wave with a spatial
extent around 5 grain diameters,  that propagates along the
direction of impact. Remarkably, the solitary wave excitation is
to a good approximation, the soliton-like solution first seen by
Nesterenko in a one-dimensional  granular chain of macroscopic
non-cohesive particles interacting via the Hertz-potential
\cite{Nesterenko_Book,Coste2,Daraio2005,Job,Sen,Nesterenko_1994}.

\subsection{Exponential decay} 

As the
solitary wave propagates through a weakly disordered hexagonal packing (see inset of Fig.\ (\ref{1D}a), 
it begins to attenuate and the initial stages of
this attenuation is well approximated as an  exponential decay.
The isotropic elasticity of the hexagonal packing allows us to
model the solitary wave as a one dimensional excitation -- we make
use of the quasi-particle approximation to understand its decay in
this regime \cite{Tichler_2013}. Consequently, we model the
interaction of the solitary wave quasi-particle with the mass
inhomogeneity as an elastic collision process (conserving momentum
and energy) that results in the disintegration of the solitary
wave into two new solitary waves during each collision.  After
propagating through $n$ grains (undergoing multiple collisions) we
find that the mean energy of the solitary wave is given by (see
App. A)
\begin{eqnarray}
\frac{\langle T_n\rangle}{T_0} \sim e^{-\frac{n}{4}\epsilon^2},
\label{theory}
\end{eqnarray}
where, $T_0$ is the initial energy of the solitary wave, $T_n$ is
the solitary wave energy after traversing $n$ beads and $\epsilon$
is the variance in the masses of the beads, assumed to be
distributed as a Gaussian random variable. As shown in the inset to
Fig.\ (\ref{1D} a), we find this estimate to be in very good
agreement with our numerical observations on an hexagonal packing
and weakly-disordered granular chains
\cite{Manju_2012}.

\subsection{Power law attenuation.} By contrast, for the hexagonal
packing with strong mass disorder Fig.\ (\ref{1D} b), the
exponential regime is barely identifiable and the solitary wave
begins to evolve into a shock like triangular profile. For the
decay in this regime, we find a striking similarity in all three
cases : the long time decay in a hexagonal packing with weak mass
disorder, the dominant regime of decay in a hexagonal packing with
strong mass disorder (large $\epsilon$) as well as the dominant
mechanism of decay in amorphous jammed packings, see Fig.\
(\ref{1D} a-c). In all cases, a triangular shock like profile
emerges, whose leading edge decays as a power law $x^{-r}$ with an
exponent approximately $r\approx 0.5$ (solid red line). This
exponent is thus independent of the amount of disorder. 

Notice, the shock like profile is obtained as an envelope over
several smaller excitations that span the system up to the leading
propagating edge. Thus, unlike the initial solitary wave
excitation whose spatial extent is around 5 grain diameters, the
shock solution spans several hundreds grains. In a recent study, the
existence of such long wavelength triangular shock like solutions
in a one dimensional chain of granular particles was established
\cite{Calvo_2012}.  Here, we find that due to material
inhomogeneities, an initial excitation in two dimensional
disordered packings, naturally evolve into long wavelength shock
solutions. As we outline below,  the power law decay follows
as a  simple consequence of energy conservation.

In the long wavelength approximation, the average strain field $\delta$
satisfies  the continuity  equation  $\delta _t +
\delta^{1/4}\delta _x=0$, where subscripts denote partial
derivatives with respect to time $t$ and space $x$ \cite{Calvo_2012}. This equation
has a similarity solution $\delta(x,t)\sim
\left(\frac{x}{t}\right)^4$. Upon integrating once with respect to
$x$ and differentiating once with respect to $t$, we obtain the
corresponding  solution for  the particle velocity field
$\phi(x,t)\sim \left(\frac{x}{t}\right)^5$. Since the energy of
the beads enclosed within the shock envelope is conserved, it
follows that $E\sim \int^{x_f}_0 dx \phi^2(x,t)\sim
\frac{x_f^{11}}{t^{10}}=$ contant, where $x_f$ is the position of
the shock front and consequently, {$t_f \sim x_f^{11/10}$}. Thus, at
the location of the front, the jump in the velocity field scales
as {$\phi(x_f,t_f) \sim  \frac{x_f^{5}}{x_f^{11/2}}\sim x_f^{-1/2}$},
yielding the exponent $\frac{1}{2}$. As shown in Fig.\ (\ref{1D} a-c),
solid red lines, we do find good agreement with the numerically determined exponent
$ r\approx 0.5$ for the decay of the shock front in the power law
regime.


\section{Emergent hydrodynamic state}  The simple model for the exponential
decay of the solitary wave suggests that at each subsequent
collision with an inhomogenity, the solitary wave splits
approximately into two solitary waves- a leading pulse (the main
degree of freedom that is damped exponentially as a result) and a
smaller solitary wave, that is either reflected or transmitted
depending upon the mass ratio. Therefore, once we allow sufficient
time for the leading solitary wave to disintegrate completely such
that a leading pulse is no longer distinguishable from the
background, we expect to reach a state comprised of several
smaller solitary waves with different energies. The solitary waves
in turn interact with each other in-elastically, (as a consequence
of the Nesterenko equation of motion being non-integrable
\cite{Nesterenko_Book}) and thus their interaction may be thought
of as inherently dissipative. However, in addition to these
processes that seek to distribute the energy initially
concentrated in a single solitary wave excitation into multiple
smaller solitary waves, the structural disorder in two dimensional
amorphous packings also spills a part of the energy into
transverse degrees of motion. Through a series of such intrinsic
dissipative mechanisms, we eventually reach a fluctuating
equilibrium-like state that spans the entire finite-sized packings under-investigation.

In order to rationalize the physics behind this fluctuating state that at first appears to be just noise, we assume that there is a well defined long wavelength, small
frequency collective hydrodynamical mode with wave-number $k$ in the $x$-direction. Upon binning in the $x$ direction, we 
define the coarse grained particle current density $\mathbf{j}(\mathbf{r},t) =
\frac{1}{\sqrt{N}}\sum^{N}_{i=1}\mathbf{v}_i(t)\delta(\mathbf{r}-\mathbf{r}_i(t))
$ and its Fourier transform $j_\alpha(\mathbf{k},t) =
\frac{1}{\sqrt{N}}\sum^{N}_{i=1}v_{i\alpha}(t)e^{i\mathbf{k}\cdot\mathbf{r}_i(t)}$.
(The bold face is a vector notation used for
spatial coordinates whose cartesian components are denoted by $\alpha$). 
Upon setting $\mathbf{k}=(k,0)$ and $\alpha=x$ or $y$, we determine numerically the 
corresponding longitudinal and transverse current
density auto-correlation functions as $C_{l,t}(k,t)=\langle
j^*_{l,t}(\mathbf{k},0)j_{l,t}(\mathbf{k},t)\rangle
\label{def-correlation}$, where the angular brackets denote
ensemble averaging over the initial time. The longitudinal and
transverse power spectral densities $P_{l,t}(k,\omega)$ are evaluated numerically using
fast Fourier transform from the respective current density
auto-correlation functions as
\begin{equation}
P_{l,t}(k,\omega)=\int^{\infty}_{-\infty}dt\ e^{i\omega
t}C_{l,t}(k,t) .
\end{equation}

\begin{center}
\begin{figure}
\includegraphics[width=0.4\textwidth]{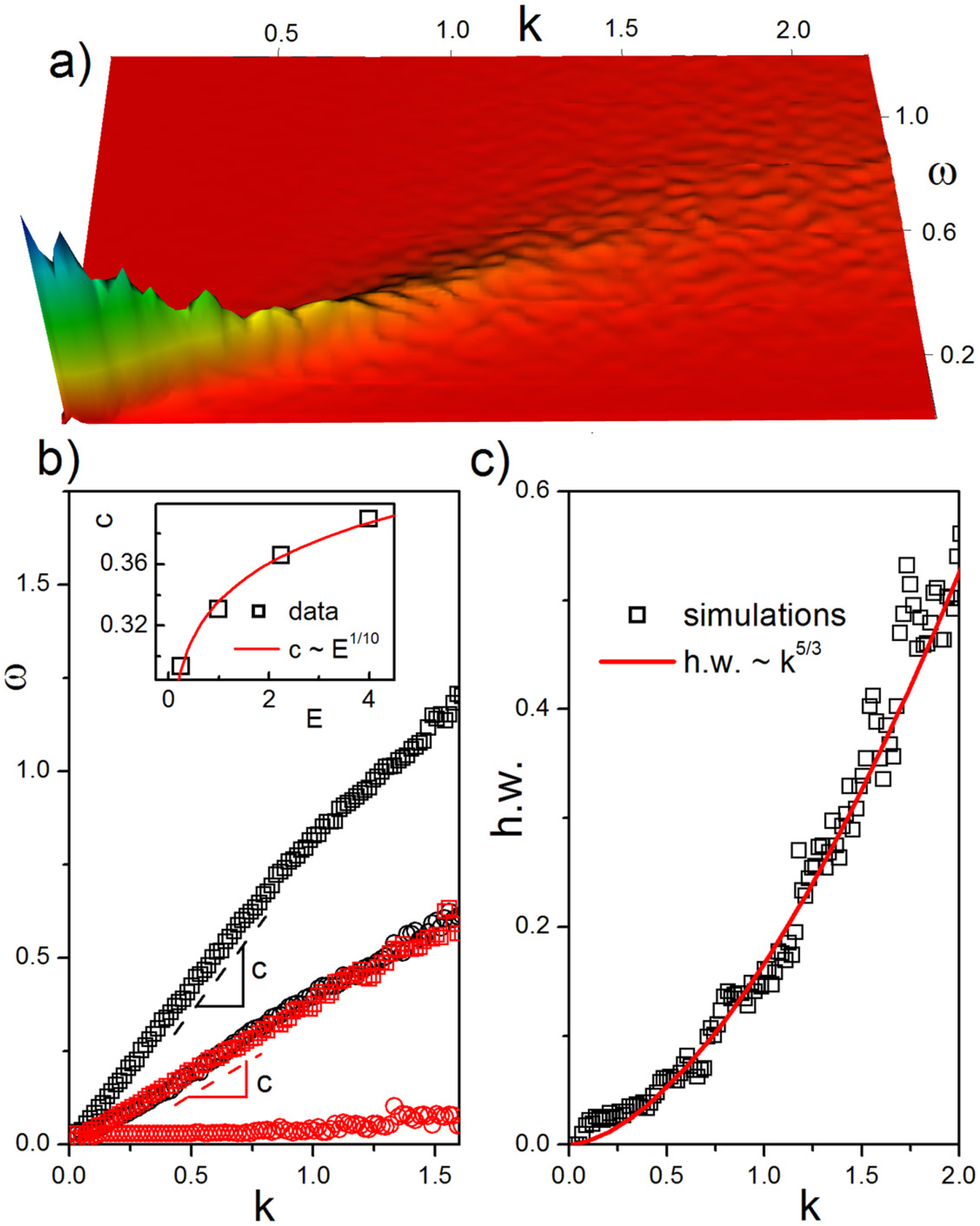}
\caption{\label{2D}(a) The power spectral density for longitudinal
modes that emerge in marginally compressed amorphous packings
($P\sim 10^{-6}$) by the complete disintegration of an initial
solitary wave excitation. (b) The red squares shows the
numerical data for the longitudinal dispersion curves.  The slope
of the linear regime scales with the energy of the solitary wave
as $c\sim E^{1/10}$ for Hertzian interaction, that we identify
with long wavelength longitudinal hydrodynamical modes.  In
contrast, the red circles shows the numerically obtained data for
the transverse dispersion curve where no linear regime is seen.
The shear mode is therefore non-propagating. For comparison, shown
are  linear dispersion curves for longitudinal (black squares) and
transverse (black circles) obtained for highly compressed jammed
packings, prepared at a pressure P$\sim 10^{-1}$.(c) The
half-width obtained numerically from the longitudinal modes as a
function of wavenumber on a linear scale and compared against the
analytical estimates h.w.$\sim k^{5/3}$ (solid red line). }
\end{figure}
\end{center}

In Fig.\ (\ref{2D} b), we plot the longitudinal (red squares)
and transverse (red circles) dispersion curves obtained
numerically  from the power spectral density Fig.\ (\ref{2D} a)
in the emergent fluctuating state.  For
comparison, we plot the corresponding longitudinal (black squares)
and transverse (black circles) dispersion curves for a highly
compressed jammed packing far from the critical density, prepared
at a pressure of P$\sim 10^{-1}$. Since the total potential energy
of a jammed packing is related to its pressure via E$\sim
\text{P}^{5/3}$, we choose an initial impact speed $U_p \approx
2.0$ to generate a solitary wave in the weakly compressed packings
(P$\sim 10^{-6}$).  This leads to a total initial solitary wave
energy $E_{SW}$ that is comparable to the energy of the highly
compressed packing $E_{PC}$ and thus allows for a more meaningful
comparison between the two states.

As seen in Fig.\ (\ref{2D} b), highly compressed jammed packings
behave as ordinary solids with a finite bulk and shear modulus and
this translates into a finite sound speed manifest in the linear
regime of the longitudinal (black squares) and shear (black circles)
dispersion curves. In contrast, exciting a jammed packing prepared
at a very small pressures (that a-priori has zero longitudinal and
shear sound speeds), leads to a linear dispersion regime for the
longitudinal modes, but no such regime is obtained for the
transverse modes.  The slope of the linear regime in the
longitudinal dispersion curves, corresponds to the speed of long
wavelength hydrodynamical sound modes. Defining the sound speed
$c$ as the second derivative of the induced potential energy;
$E_{SW}$; leads to the relation, $c \sim E_{SW}^{1/10}$
\cite{Zhirov_2011}, closely matching the numerical data in Inset
to Fig.\ (\ref{2D} b), red squares.  Thus, the speed of the long wavelength
hydrodynamical sound modes scales with the energy of the initial
solitary wave $E_{SW}$ injected into the system.

This is analogous to the scaling relation for pre-compressed
jammed packings at a finite packing fraction $\delta$, where the
sound speed scales as $c\sim E_{PC}^{1/10}$, with $E_{PC}$ being
the potential energy due to the finite pre-compression $\delta$.
Thus, in so far as longitudinal sound modes are concerned, a
rigidity induced by statically compressing a marginally compressed
packing is analogous to the rigidity induced by exciting a
marginally compressed packing with a finite energy wave. Note
therefore, one can easily replace the source of energy by a heat
bath and thereby obtain a thermally induced rigidity upon making
the substitution E$\rightarrow k_BT$. However, unlike a state that
is truly in thermal equilibrium, an external perturbation over the
fluctuating state created by the disintegration of a solitary
wave, will further raise its energy due to the absence of a
fluctuation-dissipation mechanism. The emergent state is thus at
best described as a quasi-equilibrium state.

In contrast to the longitudinal modes, the transverse modes
obtained by energizing a marginally compressed packing do not show
a well defined linear regime, see Fig.\ (\ref{2D} b), red circles.
This is in stark contrast  from a statically compressed jammed
packing where a linear transverse dispersion regime (owing to the
finite shear modulus) with a  slope that scales with the amount
of pre-compression (and does not depend upon the solitary wave
energy injected) as $c\sim E_{\text{PC}}^{1/5}$ is expected (Fig.\
(\ref{2D} b), black circles).  Thus, the shear modes excited by
injecting energy into a packing near its critical point are purely
diffusive and the medium does not develop a finite shear modulus.
There is therefore a profound difference between the resultant
states obtained by (a) statically compressing a granular packing near
its critical point versus (b) injecting energy either in the form of an
excitation or thermally. In the former case, one obtains a
solid-like medium with finite bulk and shear moduli, while in
the latter, we fluidize the medium. This means that, even after the non-linear wave excitations (characteristic of the jamming point) are strongly attenuated, one cannot
describe the system purely in terms of the normal modes of a (linear elastic) solid because the system has {\it de facto} become a fluid. This is one of the key conclusions of our work and it heralds the signal property of packings at the jamming threshold: they are fragile.  

In Fig.\ (\ref{2D} c), we show how the half width (inverse
lifetime  $\tau$ ) of the longitudinal hydrodynamical modes depend
upon the wavenumber $k$. We find that it scales anomalously as
$\tau^{-1} \sim k^{1.5}$.  For purely hydrodynamical modes obeying
the Navier-Stokes equation, the half width scales with the wave
number as hw $\sim \eta k^2$, where $\eta$ is the shear viscosity.
However, extensive numerical and analytical studies have shown
that in one dimension the time correlation functions
(whose long time integrals by definition correspond to macroscopic
transport properties such as diffusivity and shear viscosity) do
not decay exponentially but display long time tails, decaying as
power laws instead \cite{Zwanzig,Ernst,Ernst_2}.  This phenomenon
indicates the breakdown in low dimensions of the standard continuum assumption embedded in the Navier-Stokes 
equation -- the strength of fluctuations is too strong for simple coarse-grained
theories to hold.  

Analytic studies based on mode coupling
theory predict that for $d=1$ dimensions, the shear viscosity
itself scales with the wave number (in the small wave number
limit) as $\eta(k)\sim k^{-1/3}$ and thus to leading order, we obtain
a half width that scales as hw$\sim \eta(k)k^2= k^{5/3}$.  These results
have also been validated numerically in more recent studies on
one dimensional non-linear spring chains  \cite{Pikovsky,
Zhirov_2011}. The predicted $5/3$ power law dependence is consistent with our findings in Fig.\
(\ref{2D} c) solid red line. Our packings are two-dimensional but the emergent hydrodynamic description
is effectively one-dimensional because we are injecting compressional solitary wave fronts in the $x$ directions only, see {Fig. (1)}. Thus,  despite the absence of any
microscopic dissipative mechanism in our simulations, there is an emergent shear
viscosity that scales with the wave number and gives the
longitudinal and shear hydrodynamical modes a finite lifetime.

\section{Conclusion} To conclude,  we find that in response to an
impulse, hexagonal packings with mass disorder and amorphous
packings, initially generate a well defined solitary wave
excitation that is progressively attenuated. For weakly disordered hexagonal packings, the
amplitude of the solitary wave excitation decays exponentially at
early times, with a rate that depends upon the amount of disorder.
In the long time limit, we observe a transition to a power law
regime,  with a universal exponent that is independent of the
amount of disorder and is the dominant mechanism of decay in the
strongly disordered case, where an exponential decay is no longer
identifiable.  We find that the physics of the solitary wave
attenuation in jammed amorphous packings at their critical point
corresponds to the strongly disordered regime in hexagonal
packings where an initially well defined solitary wave soon
transitions into a triangular shock like profile, decaying as a
power law. We understand the two stages of decay using simple
analytical models and scaling arguments. We then study the
resultant  fluctuating state (similar to a thermalized state) that
emerges after the complete disintegration of the initial solitary
wave. This fluid-like state has emergent macroscopic properties
absent in the unperturbed granular packing: (i) a finite pressure
that scales with the energy originally injected in the form of a
solitary wave  and (ii) a finite viscosity, despite no
microscopic mechanism of dissipation being present.

\section{Acknowledgments} We acknowledge discussions with J. M. J. Leeuwen and M. van Hecke.
NU and LRG acknowledge financial support from FOM, Shell, Universidad Nacional del Sur and CONICET.

\appendix

\section{The exponential decay}
Consider  a lattice of beads of mass $m$, where nearest neighbours
interact with a one sided non-linear potential of the form
$V(\delta)=\frac{k}{\alpha}\delta^{\alpha}$, with $\delta$ being
the average overlap between neighbouring beads, $k$ the force
constant and $\alpha>2$ the exponent of the non-linear potential. If the initial overlap; $\delta\rightarrow
0$, the effective spring constant defined as the second derivative
of the potential vanishes, and as result, the lattice does not
support linear sound at zero temperature. Consequently, mechanical
strains generated in response to an impulse evolve into long-lived
non-linear solitary waves and the solitary wave energy  $E$ and
momentum $P$  satisfy the classical relation $E =  \frac{P^2}{2
m_{\text{eff}}}$, where $m_{\text{eff}}\sim 1.3m$ is the mass of
the solitary wave quasiparticle as a function of the mass of the
beads $m$ \cite{Nesterenko_1984,Job}.  Here, the momentum
$P=m_{\text{eff}}U$ where, $U$ physically corresponds to the
solitary wave amplitude.

The quasi-particle approximation of the solitary wave has been
used to study the disintegration of a solitary wave across a mass
interface \cite{Job,Tichler_2013}.  Consider  an interface between
two regions of sonic vacuum with grain masses $m_1,m_2$
respectively. For mass ratios $A=\frac{m_2}{m_1}$ close to 1, a
solitary wave initially moving with amplitude $U_0$ is seen to
split into two new solitary waves, whose amplitudes
$U_{1,f},U_{2,f}$ may be obtained using an elastic collision model
that conserves the quasi-particle energy and momentum -
\begin{subequations}
\label{collision}
\begin{eqnarray}
m_{1,\text{eff}}U_0 &=& m_{1,\text{eff}}U_{1,f} + m_{2,\text{eff}}U_{2,f}, \nonumber\\
m_{1,\text{eff}}U_0^2 &=& m_{1,\text{eff}}U_{1,f}^2 +
m_{2,\text{eff}}U_{2,f}^2, \nonumber
\end{eqnarray}
\end{subequations}

from which one can easily derive

\begin{subequations}
\label{Velocity}
\begin{eqnarray}
U_{1,f} &=& \left(\frac{1-A}{1+A}\right)U_0 , \nonumber\\
U_{2,f} &=& \left(\frac{2}{1+A}\right)U_0 \nonumber.
\end{eqnarray}
\end{subequations}

Thus, the ratio of transmitted to incident energy is

\begin{eqnarray}
\frac{m_{2,\text{eff}}U^2_{2f}}{m_{1,\text{eff}}U^2_{0}}\equiv
\frac{T_E}{T_0} = \frac{4A}{(1+A)^2} \nonumber.
\end{eqnarray}
\\
\\
In order to study the propagation of the solitary wave in a medium
with weak mass inhomogeneity,  consider a chain of beads with the
mass ratio of neighbouring beads $i,j$ related via $m_{i}=
A_{i,j}m_j$, where $A_{i,j} = 1+\text{N}_{i,j}(0,\epsilon^2)$,
The normal random variable $\text{N}_{i,j}(0,\epsilon^2)$ has
mean 0 and variance $\epsilon^2$.   Upon appealing to the
localized nature of the solitary wave (its width being around 5
bead diameters and also independent of the amplitude of the
solitary wave), we treat each bead as an interface and invoke
the quasi-particle elastic collision model. 

Thus, the energy of
the leading solitary wave after the first collision is
\begin{eqnarray}
T_1 &=& \frac{1+\epsilon\text{N}_1(0,1)}{(1+\frac{\epsilon}{2}\text{N}_1(0,1))^2}T_0 \nonumber\\
    &\sim& \left[1+\epsilon\text{N}_1(0,1))(1-\epsilon\text{N}_1(0,1) + \frac{3}{4}\epsilon^2\text{N}^2_1(0,1)\right]T_0 \nonumber\\
    &\sim& \left[1-\frac{1}{4}\epsilon^2\text{N}^2_1(0,1)\right]T_0 \nonumber,
\end{eqnarray}
where, we have retained terms up to order $\epsilon^2$. Iterating
this process $n-$ times, we find that the solitary wave energy
after it has propagated {\it n} beads diameters is
\begin{eqnarray}
T_n \sim \left[1-\frac{1}{4}\epsilon^2\text{N}^2_n(0,1))\cdot\cdot
(1-\frac{1}{4}\epsilon^2\text{N}^2_1(0,1)\right]T_0 \nonumber.
\end{eqnarray}
Retaining terms only to order $\epsilon^2$, we find
\begin{eqnarray}
\frac{T_n}{T_0} &\sim& 1- \frac{1}{4}\epsilon^2\sum^n_{k=1}\text{N}^2_k(0,1), \nonumber\\
        &\sim& 1- \frac{1}{4}\epsilon^2\chi^2(n) \nonumber,
\end{eqnarray}
where, $\chi^2(n)$ is the chi function with an expectation value
$n$. Taking the mean, we obtain that the average solitary wave
energy after propagating $n$  beads diameters reads
\begin{eqnarray}
\frac{\langle T_n\rangle}{T_0} \sim& 1- \frac{1}{4}n\epsilon^2
\sim e^{-\frac{n}{4}\epsilon^2}. \label{theory}
\end{eqnarray}
This result agrees quantitatively with numerical simulations in 1D \cite{Manju_2012} and qualitatively with our observations in 2D packings.

%

\end{document}